\documentclass[10pt,conference]{IEEEtran}
\IEEEoverridecommandlockouts

\usepackage{algorithm,amsbsy,amsmath,amssymb,epsfig,bbm,mathrsfs,multirow,amsthm}
\usepackage{color}
\usepackage{algpseudocode}
\usepackage{graphicx, caption, subcaption, array, multirow, bm}
\usepackage{makecell}
\DeclareMathAlphabet{\mathpzc}{OT1}{pzc}{m}{it}

\usepackage{cases}

\usepackage{helvet}
\usepackage{subcaption}  

\usepackage[a4paper, top=0.75in, left=1in, right=0.662in, bottom=1in, footskip=0.5in]{geometry}
\def\bq{{\bf q}}

\def\bq{{\bf q}}

\newtheorem{definition}{\bf Definition}

\def\BibTeX{{\rm B\kern-.05em{\sc i\kern-.025em b}\kern-.08em     T\kern-.1667em\lower.7ex\hbox{E}\kern-.125emX}}
\begin{document}
	

\title{Resilience Optimization in 6G and Beyond Integrated Satellite-Terrestrial Networks: A Deep Reinforcement Learning Approach} 
\author{
	\IEEEauthorblockN{Dinh-Hieu Tran, Nguyen Van Huynh, Van Nhan Vo, Madyan Alsenwi, Eva Lagunas,  Symeon Chatzinotas }  }

\author{\IEEEauthorblockN{1\textsuperscript{st} Dinh-Hieu Tran}
\IEEEauthorblockA{\textit{Snt, University of Luxembourg, Luxembourg}\\
hieu.tran-dinh@uni.lu}
\and
\IEEEauthorblockN{2\textsuperscript{nd} Nguyen Van Huynh}
\IEEEauthorblockA{\textit{ University of Liverpool}\\
huynh.nguyen@liverpool.ac.uk}
\and
\IEEEauthorblockN{4\textsuperscript{th} Van~Nhan~Vo}
\IEEEauthorblockA{\textit{Duy Tan University}\\
vonhanvan@dtu.edu.vn}
\and
\IEEEauthorblockN{5\textsuperscript{th} Madyan Alsenwi}
\IEEEauthorblockA{	\textit{SnT, University of Luxembourg, Luxembourg}\\
madyan.alsenwi@uni.lu}
\and
\IEEEauthorblockN{5\textsuperscript{th} Eva Lagunas}
\IEEEauthorblockA{	\textit{SnT, University of Luxembourg, Luxembourg}\\
eva.lagunas@uni.lu}
\and
\IEEEauthorblockN{6\textsuperscript{th}Symeon Chatzinotas}
\IEEEauthorblockA{	\textit{SnT, University of Luxembourg, Luxembourg}\\
symeon.chatzinotas@uni.lu}
}

\maketitle

\begin{abstract}
	Ensuring network resilience in 6G and beyond is essential to maintain service continuity during base station (BS) outages due to failures, disasters, attacks, or energy-saving operations. This paper proposes a novel resilience optimization framework for integrated satellite-terrestrial networks (ISTNs), leveraging low Earth orbit (LEO) satellites to assist users when terrestrial BSs are unavailable. Specifically, we develop a realistic multi-cell model incorporating user association, antenna downtilt adaptation, power control, heterogeneous traffic demands, and dynamic user distribution. The objective is to maximize of the total user rate in the considered area by optimizing the BS's antenna tilt, transmission power, user association to neighboring BS or to a LEO satellite with a minimum number of successfully served user satisfaction constraint, defined by rate and Reference Signal Received Power (RSRP) requirements. To solve the non-convex, NP-hard problem, we design a deep Q-network (DQN)-based algorithm to learn network dynamics to maximize throughput while minimizing LEO satellite usage, thereby limiting reliance on links with longer propagation delays and prolonging satellite operational lifetime. Simulation results confirm that our approach significantly outperforms the benchmark one.
\end{abstract}

\begin{IEEEkeywords}
	6G, Deep Q-Network, Integrated Satellite-Terestrial Networks, Network Resilience, Network Energy Saving, Self-Organizing Networks (SONs).
\end{IEEEkeywords}

\IEEEpeerreviewmaketitle

\section{Introduction}
\label{sec:Intro}


Network resilience can be defined as the ability of a network to provide a minimum level of service to users based on the existing infrastructure in the unexpected events such as failures, disasters, and attacks \cite{ITUResilience}. Network resilience becomes more urgent and difficult with the rapid development of new technologies for 6G and beyond networks such as virtual reality \cite{chen2025qoe}, energy-efficient communications \cite{tran2020coarse}, holographic \cite{Gupta2025},  metaverse \cite{qian2026enhancing}, emergency communications \cite{tran2021uav}, 3D video streaming \cite{badnava2025neural}, etc. due to the high data rate and latency requirements of these applications.

Therefore, network resilience has attracted significant attention in recent studies \cite{weedage2024resilience,tran2025network,tchimwa2024enhancing,liyuan2024hybrid,iwamoto2023deep}. In \cite{weedage2024resilience}, Weedage et al. investigated cellular network resilience by incorporating public data from different regions in Netherland. Their study showed that roaming agreements between difference mobile network operators can significantly improve resilience. In \cite{tran2025network}, Tran et al. focused on resilience in energy-saving scenarios, where BSs enter sleep mode during low traffic to reduce power consumption and CO2 emissions. In \cite{tchimwa2024enhancing}, Tchimwa et al. explored how varying traffic demands affect resilience, especially under cascading failures, and proposed a mathematical framework to assess their impact. In \cite{liyuan2024hybrid}, Lyuan et al. introduced an outage detection algorithm using a hybrid Generative Adversarial Network and an overlap-aware Artificial Neural Network to address data imbalance and class overlap, improving classification via weighted samples. In \cite{iwamoto2023deep}, Iwamoto et al. studied cell outage compensation by proposing an antenna selection algorithm based on spatial relationships and used DRL to find optimal solutions. All the above studies focus solely on terrestrial networks (TNs) for resilience support. However, NTNs have been identified as key components to complement TN coverage and ensure robust connectivity \cite{Ntontin2025}. To the best of our knowledge, only one recent study \cite{PijnappelCelloutage24} considers the use of UAVs for BS outage support, where Pijnappel et al. employed drones as flying BSs in disaster scenarios to complement traditional TN outage compensation mechanisms that adjust antenna tilts of neighboring cells. 
 
 Although previous studies have achieved certain achievements, network resilience is still receiving much attention from both academia and industry for the development of 6G and beyond networks \cite{khaloopour2024resilience, dinh2025uav}. Motivated by the above observations, we have developed this research to ensure resilience in 6G and beyond ISTNs. The main contributions in this paper will be given as follows:


\begin{itemize}
	\item To the best of our knowledge, this is the first work that investigates the network resilience in integrated satellite-terrestrial networks applying deep reinforcement learning (DRL). Given the dynamic nature of the environment and the mathematical complexity of the problem, DRL is adopted as a suitable solution approach.
	\item We propose a novel system model that accounts for various real-world challenges. Specifically, we consider multi-cell radio access networks, in which BSs turned-off/outage due to attacks, disasters or failures (i.e., time domain); neighboring BS's antenna tilt adaptation (i.e., antenna/space domain); surrounding BS's power adaptation (i.e., power domain); randomly varying users traffic requirements; and random user's location distribution. Especially, we also consider the LEO satellite (LEOS) systems to serve users who cannot be accommodated by the terrestrial BS due to overloaded or restricted resources. 
	\item Based on the proposed system model, our objective is to maximize the total throughput of all users within the coverage area while minimizing the total number of users served by LEO satellites (LEOSs), thereby prolonging their operational lifetime and minimizing the impact of transmission latency caused by long propagation delays on satellite links. This not only enhances the resilience of the terrestrial network by effectively leveraging satellite backup in the face of base station outages, but also ensures the long-term sustainability and resource efficiency of the LEOS segment.
\end{itemize}


\section{System Model}
\label{sec:Sys}

This paper considers an integrated satellite-terrestrial multi-cell downlink (DL) system, which encompasses two segments, i.e., the LEOS segment and the terrestrial network (TN). Moreover, the sets of $S$ LEOSs, $M$ gNBs, and $U$ ground users (GUs) are denoted by ${\mathcal{S}} = \{1, \dots,s, \dots, S\}$, ${\mathcal{M}} = \{1, \dots,m, \dots, M\}$, and ${\mathcal{U}} = \{1, \dots,u, \dots, U\}$, respectively. In addition, each gNB is equipped with three distinct antennas to cover three cells/sectors, where the set of antenna indexed by $i \in {\mathcal{I}} = \{1,2,3 \}$. 

\subsection{Channel Model}
The distance from gNB $m$ $\rightarrow$ GU $u$ and from LEOS $s$ $\rightarrow$ GU $u$ are respectively expressed as \cite{3gpp2019ntn}:
\begin{align}
	\label{eq:2}
	d_{um} &= \sqrt{(h_m-h_u)^2 + || \bq_m - \bq_u ||^2}, \\
	d_{us} &= \sqrt{R_{\rm E}^2 \sin^2(\beta_{us}) + h_s^2 + 2 h_{s} R_{\rm E}} - R_{\rm E} \sin(\beta_{us}),
\end{align}
where $h_m$ and $h_u$ denote the altitude of the gNB $m$ and GU $u$, respectively. Furthermore, the locations of gNB $m$ and GU $u$ are defined as $\boldsymbol{q}_j \in {\mathbb{R}}^{2 \times 1}$ or $\boldsymbol{q}=[x_l,y_l]$, where $l \in \{m,u\}$. $R_{\rm E} \approx $ 6378 km denotes the Earth radius. $h_s$ presents the LEOS altitude. $\beta_{us}$ is the elevation angle (EA) from LEOS $s$ $\rightarrow$ GU $u$.

As reported from 3GPPTR 38.881 that, the line-of-sight (LoS) occurs most of the time in outdoor environments for the satellite \cite{3gpp2019ntn,kim2024feasibility}. Therefore, the propagation model from LEOS $s$ $\rightarrow$ GU $u$ apply the non-Shadowed Rician channel model. Specifically, the downlink channel coefficient from LEOS $s$ $\rightarrow$ GU $u$ is modeled as:
\begin{align}
	\label{eq:channel_power}
	h_{us} = \sqrt{\omega_{us}} \tilde{h}_{us},
\end{align}
where $\omega_{us}$ and $\tilde{h}_{us}$ denote large scale-fading and small-scale Rician fading (SCRF) coefficients, respectively. More specifically, $\omega_{us}$ can be given as \cite{kim2024feasibility, rahman2025joint}:
\begin{align}
	\label{eq:PL}
	{\mathcal{L}}_{us} = 32.45 + 20 \log_{10}(f_c) + 20 \log_{10}(d_{us}), 
\end{align}
where $f_c$ represents the carrier frequency transmitted by the LEOS $s$, $d_{us}$ is the distance between LEOS $s$ $\rightarrow$ GU $u$.

Then, the received power at the GU $u$ from LEOS $s$ is expressed as:
\begin{align}
	\label{eq:Px}
	{\mathcal P}^r_{us} = P^t_{us} |h_{us}|^2  10^{-{\mathcal{L}}_{us}/10} G_i(\alpha_{us}, \beta_{us}) G_u ,
\end{align}
where $P^t_{us}$ is the transmit power (TxP) from satellite $s$ $\rightarrow$ GU $u$. $G_i(\alpha_{us}, \beta_{us})$ is the transmit antenna gain (ATG) of the LEOS $s$ and $G_u$ is the received ATG of GU $u$.

In the TN, we consider a realistic transmission model that accounts for line-of-sight (LoS) and non-line-of-sight (NLoS) links adapting to various environments such as urban or rural areas. Furthermore, the model incorporates large-scale path loss and SCRF, and ATGs. Accordingly, the received power at the GU $u$ from antenna $i$ of gNB $m$ is expressed as: 

\begin{align}
	\label{eq:9}
	{\mathcal P}^{i,r}_{um} = {\mathcal P}^{i,t}_k  |h_{um}^i|^2  (d_{uk}^{\rm 3D})^{-\alpha} {\mathfrak B_k^i( \alpha_{um},\beta_{um})}  G_u  , 
\end{align}
where ${\mathcal P}^{i,t}_k$ denotes the TxP of gNB $m$ through antenna $i$, $d_{um}^{\rm 3D}$ is the distance from gNB $m$ $\rightarrow$ the GU $u$, $h_{um}^i$ is the SCRF, $G_u$ denotes the received ATG at GU $u$, while ${\mathfrak B_k^i(\alpha_{um},\beta_{um})}$ represents the transmit ATG of the gNB $k$ through antenna $i$, which is defined in the antenna model subsection \ref{subsec_Antenna}.

Then, we introduce the equation to calculate the RSRP as follows \cite{RSRP}:
\begin{align}
	\label{RSRP}
	{\tilde{\mathcal P}_{us}} &= {\mathcal P}^r_{us} - 10\log{(12* \mathcal{N})}, \\
	{\tilde{\mathcal P}_{um}^i} &= {\mathcal P}^{i,r}_{um} - 10\log{(12* \mathcal{N})}, 
\end{align}
where ${\mathcal P}^r_{us}$ and ${\mathcal P}_{um}^{i,r}$ denote the Received Signal Strength Indicator (RSSI) from the LEOS $s$ and the gNB $k$, respectively. The RSSI represents the total received power at the UE, which includes the main received signal power, co-channel interference, and white noise. $\mathcal{N}$ denotes the number of resource blocks (RBs) corresponding to the given channel bandwidth.

From \eqref{eq:9}, the rate (in Bits/s/Hz) at GU $u$ from LEOS $s$ and gNB $m$ are given respectively as follows:
\begin{align}
	\label{eq:throughput}
	{\mathcal R}_{us} &=   \log_2 \left(1+\zeta_{us}\right), \\
	{\mathcal R}_{um}^i &=   \log_2 \left(1+\zeta_{um}\right),
\end{align}
with:
\begin{align}
	\label{eq:SINR_NTN}
	\zeta_{us} &\triangleq \frac{ {\mathcal P}^t_{us}  |h_{us}|^2  10^{-{\mathcal{L}}_{us}/10} {G_s(\alpha_{us},\beta_{us})}  G_u } { \big(\phi^{\rm RIC} \sum\limits_{s^\ast \in {\cal S} \setminus s} {\mathcal P}_{us^\ast}^t +  \sigma^2\big) }, \\
	\zeta_{um} &\triangleq \frac{ {\mathcal P}_{um}^{i,t}  |h_{um}^i|^2  (d_{um}^{\rm 3D})^{-\alpha} {\mathfrak B_k^i(\alpha_{um},\beta_{um})}  G_u } { \big(\phi^{\rm RIC} \sum\limits_{k^\ast \in {\cal K} \setminus k} {\mathcal P}_{um^\ast}^{i,t} +  \sigma^2\big) },
	\label{eq:SINR_TN}
\end{align}
where $\Phi^{\rm RIC} \triangleq \phi^{\rm RIC} \sum\limits_{s^\ast \in {\cal S} \setminus s} {\mathcal P}_{us^\ast}^t$ and $\Phi^{\rm RIC} \triangleq \phi^{\rm RIC} \sum\limits_{m^\ast \in {\cal M} \setminus m} {\mathcal P}_{um^\ast}^i$ represent the residual interference after canceling interference from other LEOS $s^\ast \in {\mathcal{S}} \setminus s$ and other gNB $m^\ast \in {\mathcal{M}} \setminus m$, respectively. In addition, we assume that the gNB and LEOSs operate at different frequency band \cite{rahman2025joint} and GU $u$ only connect to a gNB or a LEOS at a time. therefore, the interference is presented  as in denominators of Eqs. \eqref{eq:SINR_NTN} and \eqref{eq:SINR_TN}. $\sigma^2$ denotes the variance of the additive white Gaussian noise (AWGN) $n_0$, where $n_0 \sim {\cal{CN}}(0,\sigma^2)$. 

Since $h_{us}$ and $h_{um}^i$ are random variables (RVs), the achievable rates ${\mathcal R}_{us}$ and ${\mathcal R}_{um}^i$ are also RVs. Thus, we aim to derive approximate expressions for the achievable rates \cite{tran2021uav,HieuTVT22}, i.e.,:
\begin{align}
	\label{eq:Lemma1_1}
	\bar{\mathcal R}_{us} &= \log_2 \Bigg(1+\frac{e^{-E} \widehat{\mathcal P}_{us}   } {\nu_{us} }\Bigg), \\
	\bar{\mathcal R}_{um}^i &= \log_2 \Bigg(1+\frac{e^{-E} \widehat{\mathcal P}_{um}^i   } {\nu_{um}^i }\Bigg),
	\label{eq:Lemma1_2}
\end{align}
where $\widehat{\mathcal P}_{us} = {\mathcal P}_{us}  10^{-{\mathcal{L}}_{us}/10}  {G_s(\alpha_{us},\beta_{us})} G_u  $, $\widehat{\mathcal P}_{um}^i = {\mathcal P}_{um}^i (d_{um}^{\rm 3D})^{-\alpha}  {\mathfrak B_k^i(\alpha_{um},\beta_{um})} G_u  $, $\nu_{us} = \big(\phi^{\rm RIC} \sum\limits_{s^\ast \in {\cal S} \setminus s} {\mathcal P}_{us^\ast}^t+  \sigma^2\big)$, and $\nu_{um}^i = \big(\phi^{\rm RIC} \sum\limits_{k^\ast \in {\cal K} \setminus k} {\mathcal P}_{um^\ast}^{i,t}+  \sigma^2\big)$, $E$ equals 0.5772156649 \cite{tran2022satellite}.

\subsection{Antenna Model}
\label{subsec_Antenna}
This work adopts a practical 3D antenna pattern as specified by 3GPP \cite{TR36814}. Specifically, we first compute the EA $\alpha$ and azimuth angle (AA) $\beta$ from gNB $m$ $\rightarrow$ GU $u$ in the vertical (elevation) and horizontal (azimuth) planes, respectively, as follows \cite{TR36814}:
\begin{align}
	\label{eq:angle}
	\beta_{um} & = {\rm arctan} \left( \frac{h_k-h_u}{d_{um}^{\rm 2D}} \right), \\
	\alpha_{um} &= {\rm arctan} \left( \beta_u - \beta_m \right),
\end{align}
where $\beta_u$ and $\beta_m$ represent the AAs measured from the x-axis $\rightarrow$ the main lobe's center of gNB $k$ and from the x-axis $\rightarrow$ GU $u$, respectively. Additionally, the values of $\beta_m$ are $0^0$, $120^0$, and $-120^0$ corresponding to 3 antennas in 3 sectors.

The azimuth and elevation gain models are defined respectively as follows \cite{TR36814}:
\begin{align}
	\label{azimuthgain}
	{\mathfrak B_k^i(\alpha_{um}) }= -\min \left\{12 \left(\frac{\alpha_{um}}{\alpha_{\rm 3dB}} \right)^2, {\mathcal F} \right\} + {\mathfrak B_{\max}}, \\ \label{elevationgain}
	{\mathfrak B_k^i(\beta_{um}) }= -\min \left\{12 \left(\frac{\beta_{um}- \beta_{um}^{\rm tilt} }{\beta_{\rm 3dB}} \right)^2, {\mathpzc S }\right\} ,
\end{align}
where ${\mathcal F} = 20$ dB represents the front-back ratio; ${\mathfrak B_{\max}}$ = 14 dBi denotes the maximum ATG; $\alpha_{\rm 3dB} = 70^0$ reprresents the azimuth half-power beamwidth; The antenna down-tilt angle $\beta_{um}^{\rm tilt}$ comprises both mechanical (i.e.,$\beta_{um}^{\rm mtilt}$) and electrical (i.e.,$\beta_{um}^{\rm etilt}$) components. Variations in this down-tilt angle affect both $\beta_{um}$ and $\alpha_{um}$ values;  The azimuth and elevation half-power beamwidths, denoted by $\beta_{\rm 3dB}$ and $\beta_{\rm 3dB}$, can both be set to $65^0$; 

To characterize the full 3D ATG pattern, these two horizontal and vertical components need to combine \cite{TR36814} as follows:
\begin{align}
	\label{3Dgain}
	{\mathfrak B_k^i(\beta_{um}, \alpha_{um})} = {\mathfrak B_k^i(\beta_{um})} + {\mathfrak B_k^i(\alpha_{um})}.
\end{align}

\section{Integrated Satellite-Terrestrial for Network Resilience Problem}
\label{sec:Problem}
This section describes the formulation of the resilience-oriented integrated satellite-terrestrial optimization problem. In the considered system, when one or more gNBs are unavailable, the neighboring gNBs and coverage LEOSs serve the GUs previously covered by the inactive nodes. However, due to limited available resources, not all GUs can be guaranteed their required QoS, such as rate or RSRP. 

A gNB $m$ may be deactivated due to outage or energy-saving considerations, leading to:
\begin{subnumcases} 
	{\psi_{m}  =}
	1, \hfill \text{gNB} \; m \; \text{is active}, \label{eq:RSRPa}\\
	0, \hfill \text{gNB} \; m \; \text{is off}.  \label{eq:RSRPb}
\end{subnumcases}

The GUs can be served by either satellite $s$ or gNB $m$, provided that its RSRP is satisfied, we have the following user association constraints:
\begin{subnumcases} 
	{\kappa_{um}  =}
	1, \hfill {\tilde{\mathcal P}_{um}^i} \ge {\tilde{\mathcal P}_{um}}^{\rm th} \label{eq:RSRPa},\\
	0, \hfill {\tilde{\mathcal P}_{um}^i} \le {\tilde{\mathcal P}_{um}}^{\rm th}. \label{eq:RSRPb}
\end{subnumcases}
\begin{subnumcases} 
	{\kappa_{us}  =}
	1, \hfill {\tilde{\mathcal P}_{us}} \ge {\tilde{\mathcal P}_{us}}^{\rm th} \label{eq:RSRPNTNa},\\
	0, \hfill {\tilde{\mathcal P}_{us}} \le {\tilde{\mathcal P}_{us}}^{\rm th}. \label{eq:RSRPNTNb} 
\end{subnumcases}
\begin{align}
		\sum\limits_{s \in {\cal S}, m \in {\cal M} } (\kappa_{um} + \kappa_{us} ) \leq 1, 
\end{align}

If GU $u$ meets the RSRP condition, an additional constraint is proposed to ensure that its rate requirement is also satisfied, i.e.,:
\begin{subnumcases} 
	{\zeta_{um}  =}
	1, \hfill {\mathcal R}_{um}^i \ge {\mathcal R}_{u}^{\rm th} \label{eq:RSRPa}\\
	0, \hfill {\mathcal R}_{um}^i \le {\mathcal R}_{u}^{\rm th}  \label{eq:RSRPb}
\end{subnumcases}

\begin{definition}
	GU \( u \) is considered successfully served if and only if both its rate and RSRP requirements are satisfied.
\end{definition} 

Then, we introduce a new binary variable $\pi$ such that:
\begin{subnumcases} 
	{\pi_{uk}  =}
	1, \hfill \kappa_{uk}   = 1 \; $\&$ \; \zeta_{um}  = 1,\label{eq:RSRPa}\\
	0, \hfill {\rm Otherwise}. \label{eq:RSRPb}
\end{subnumcases}
with $k \in \{m,s\}$.

Let us define ${\boldsymbol \kappa} \triangleq \{\kappa_{uk}, \forall u,k \in \{m,s\}\}$, ${\boldsymbol \zeta} \triangleq \{\zeta_{uk}, \forall u, k \in \{m,s\} \},$  ${ \boldsymbol \pi} \triangleq \{{\pi_{uk}}, u, k \in \{m,s\} \}$, ${ \boldsymbol \beta} \triangleq \{\beta_{um}^{\rm tilt}, m \in {\cal M} , u \in {\cal U}\}$, and ${ \boldsymbol {\mathcal P}_m^i} \triangleq {\mathcal P}_{um}^{i,t}, \forall u,m$. Building on the above discussion, the objective is to maximize the total achievable throughput of all GUs with minimum number of served LEOSs, subject to user association, RSRP and QoS constraints, while ensuring that a specified number of GUs are successfully served. The optimization problem is formulated as follows:

\begin{align}
	\label{eq:P1}
	{\cal P}_1:\ &\max_{{\boldsymbol \kappa}, {\boldsymbol \zeta}, { \boldsymbol \pi}, {\boldsymbol \beta}, { \boldsymbol {\mathcal P}}_m^i }~~ \sum\limits_{m \in {\cal M}, u \in {\cal U}} \pi_{um}  \bar{\mathcal R}_{um}^i \\ \notag &+ \sum\limits_{s \in {\cal S}, u \in {\cal U} }   \pi_{us}  \bar{\mathcal R}_{us} - \lambda \sum\limits_{s \in {\cal S}, u \in {\cal U} }  \pi_{us}   \\
		\textbf{s.t.:} \notag \\
	& C1: \{\psi_{m}, \kappa_{um}, \kappa_{us}, \zeta_{um}, \pi_{um} \} \in \{0,1\}, \forall u,m,s \IEEEyessubnumber\label{eq:P1:b}\\
	&C2: \sum\limits_{s \in {\cal S}, m \in {\cal M} } (\kappa_{um} + \kappa_{us} ) \leq 1, \\
		&  C3: \log_2 \Bigg(1+\frac{e^{-E} \widehat{\mathcal P}_{uk} }{ \nu_{uk} }\Bigg) \ge {\mathcal R}_{uk}^{\rm th}, \forall u,k \in \{m,s\} \IEEEyessubnumber\label{eq:P1:d}\\
	&  C4: \left\|  {\boldsymbol \pi}  \right\|_1  \ge {\boldsymbol \pi}^{\min}, \IEEEyessubnumber\label{eq:P1:e}\\
		& C5: \sum\limits_{ u \in {\cal U} } {\pi}_{uk} \le \pi_{k}^{\max}, \forall u, k \in \{m,s\} \IEEEyessubnumber\label{eq:P1:f}\\
	& C6: {\mathcal R}_{um}^{\min} \le {\mathcal R}_{um}^{\rm th} \le {\mathcal R}_{um}^{\max}, \forall u,m, \IEEEyessubnumber\label{eq:P1:g}\\
	& C7: {\mathcal P}_{m}^{\min} \le {\mathcal P}_{um}^{i,t} \le {\mathcal P}_{m}^{\max}, \forall u,m, \IEEEyessubnumber\label{eq:P1:h}\\
	& C8: \beta_{um}^{\rm tilt} \in \left[0^\circ,14^\circ\right], \forall u,m. \IEEEyessubnumber\label{eq:P1:k}
\end{align}
where the first and the second component in \eqref{eq:P1} represent the total achievable throughput from gNBs and LEOSs; The third component in \eqref{eq:P1} is a penalty term to minimize the number of usage LEOS to prolong LEOs lifetime. More over, the goal of minimizing the number of LEOs is to ensure continuous service while balancing between overall system resilience and long-term sustainability which aligns with 6G network resilience objectives. $\lambda$ is the penalty coefficient; C2 represents the maximum number of served gNB/LEOS to one GU $u$ is less than or equal to 1; C3 is the rate requirements of GU $u$ when it is served by gNB $m$ or LEOS $s$; C4 denotes we need to serve at least a minimum number of GUs; C5 presents the maximum number of successfully served GUs by one gNB/LEOS, showing the limited resources of the gNB and LEOS; C6 is the random traffic demand of GUs; C7 shows the limited transmit power of gNB $m$; C8 is the limited antenna tilt of a gNB.

Problem $ \mathcal{P}_1 $ is a NP-hard and non-convex problem. We then propose an efficient solution approach based on DQN in the next section.

\section{Deep Q-Network-based Solution}
\label{sec:DQN}
To address the challenges arising from BS outages, heterogeneous traffic demands, and user distribution, we formulate the problem as a Markov Decision Process (MDP). The MDP is defined by a tuple $(\mathcal{S}, \mathcal{A}, \mathcal{R})$, where $\mathcal{S}$ denotes the state space, $\mathcal{A}$ the action space, and $\mathcal{R}$ the reward function. We develop a DQN-based learning algorithm to dynamically adapt neighboring BS antenna tilt, and transmit power.

\subsection{State Space}
The state space captures the configuration of antenna downtilt angles and transmit power levels for each gNB sector. Specifically, each gNB comprises three sectors, and each sector is characterized by a pair $( {\mathcal P}_{um}^{i,t}, \beta_{um}^{\rm tilt} )$ where ${\mathcal P}_{um}^{i,t}$ is the transmit power and $\beta_{um}^{\rm tilt}$ is the downtilt angle. Thus, the state for a gNB is defined as a matrix $\mathbf{S} \in \mathbb{R}^{L \times 2}$, where $L$ denotes the number of cell/sector per gNB.

\subsection{Action Space}
In each sector, the DQN agent selects from three discrete downtilt adjustments: $\{-1^\circ, 0^\circ, +1^\circ\}$, and three discrete power level adjustments: $\{-5\,\mathrm{dB}, 0\,\mathrm{dB}, +5\,\mathrm{dB}\}$. This results in 9 possible combinations per sector, and $9^L $ composite actions per gNB (agent).

\subsection{Reward Function}
The reward function encourages the maximization of network throughput while discouraging reliance on LEOS backup. The instantaneous reward is given by:
\begin{subnumcases}
	{{\mathcal R}_{s,a,s'}  =}
	\sum\limits_{m \in {\cal M}, u \in {\cal U}} \pi_{um}  \bar{\mathcal R}_{um}^i + \sum\limits_{s \in {\cal S}, u \in {\cal U} }   \pi_{us}  \bar{\mathcal R}_{us} \\ \notag - \lambda \sum\limits_{s \in {\cal S}, u \in {\cal U} }  \pi_{us} \hfill {\text{If} } \hfill \pi_{uk}   = 1, \label{eq:rewarda}\\
	0, \hfill {\text{Otherwise} }. \label{eq:rewardb}
\end{subnumcases}
where $\lambda$ is a penalty coefficient.

%
%

\subsection{Deep Q-Network}
In the Deep Q-Network (DQN) approach, we employ a deep neural model $\mathcal{Q}_\theta(\mathbf{s}, \mathbf{a})$ to approximate the optimal action-value function $\mathcal{Q}^*(\mathbf{s}, \mathbf{a})$, where $\theta$ denotes the trainable parameters of the neural network. The input to the network is the current state $\mathbf{s}$, and the output is a vector of $\mathcal{Q}$-values corresponding to all possible actions. The goal is to minimize the discrepancy between predicted and target $\mathcal{Q}$-values.

To enhance learning stability and efficiency, we adopt an experience replay strategy. Interactions with the environment are stored as transitions $\tau_i = (\mathbf{s}_i, \mathbf{a}_i, r_i, \mathbf{s}_{i+1})$ in a memory buffer $\mathbb{M} = \{\tau_1, \dots, \tau_N\}$. 

During training, a mini-batch of samples is randomly drawn from $\mathbb{M}$ to update the DNN parameters using the following temporal-difference update:
\begin{equation}
	\mathcal{Q}_\theta(\mathbf{s}_i, \mathbf{a}_i) \leftarrow r_i + \gamma \cdot \max_{\mathbf{a}'} \mathcal{Q}_{\theta'}(\mathbf{s}_{i+1}, \mathbf{a}'),
\end{equation}
where $\gamma$ is the discount factor $(0 < \gamma \leq 1)$, and $\theta'$ represents the parameters of a separate target network that is updated periodically to stabilize training.

Action selection follows an $\varepsilon$-greedy strategy. Initially, the agent explores the action space with high probability ($\varepsilon = 1$), which is gradually reduced to a minimum value ($\varepsilon = 0.01$) to promote exploitation of the learned policy. This balances exploration of new actions with exploitation of known rewarding actions over time.

\begin{figure}[t!]
	\centering	\includegraphics[scale=0.45]{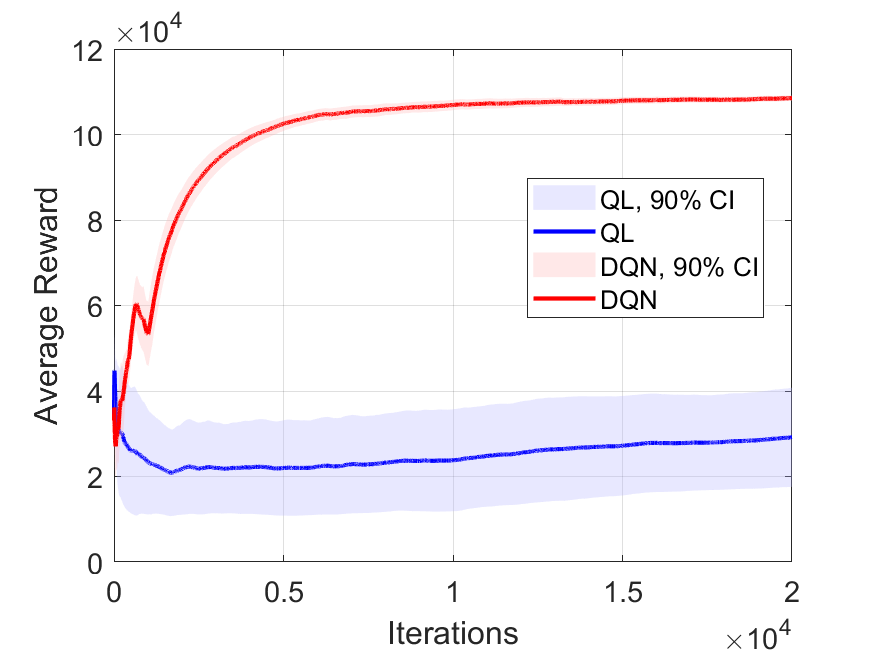}
	\caption{Average rewards vs. number of iterations.}
	\label{Fig3}
\end{figure}

\begin{figure*}[t]
	\centering   
	\begin{subfigure}[b]{0.4\textwidth}
		\includegraphics[width=\textwidth]{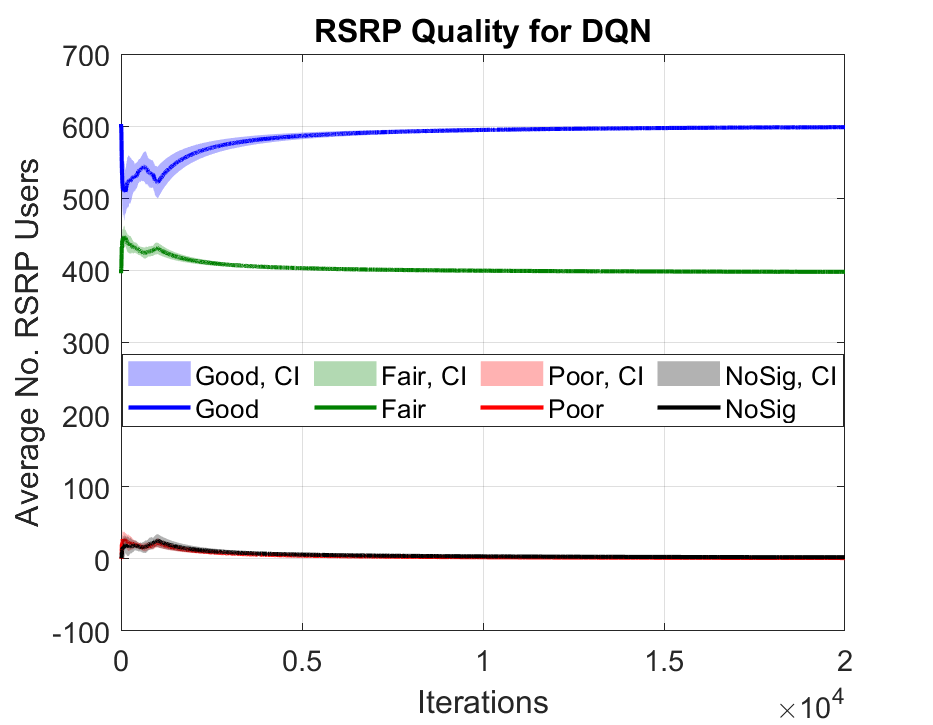}
		\caption{Average RSRP Users for DQNA}
		\label{fig:4a}
	\end{subfigure}
	\begin{subfigure}[b]{0.4\textwidth}
		\includegraphics[width=\textwidth]{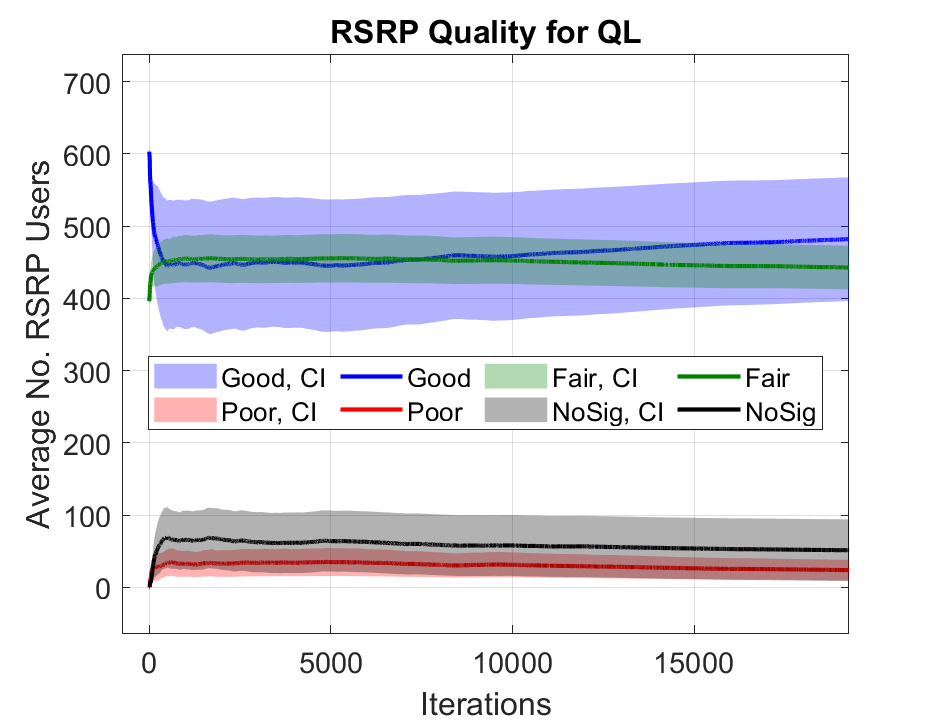}
		\caption{Percentage RSRP Users for QLA}
		\label{fig:4b}
	\end{subfigure}
	\caption{Average number of RSRP Users vs. Iterations.}
	\label{Fig4}
\end{figure*}
\section{Simulation Results}
\label{Results}
This section presents simulation results to evaluate the proposed joint deactivated gNBs, RSRP constraint, user's QoS requirement, as well as the adapted antenna tilt and power adaptation of surrounding gNBs to cover the outage one in 6G and beyond networks. We consider a system with $U$ users that are randomly distributed in $M$ gNBs following a uniform distribution. Moreover, these users demand different QoS requirements, i.e., data rates. The simulation parameters are set as follows: the number of users $|\mathcal{U}|=1000$, the height of the user is $h_u=1.5$ meters, and the height of the gNB is $h_k=10$ meters. The down-tilt value is $\beta_{um}^{\mathrm{tilt}} \in [0^\circ, 14^\circ]$. The minimum and maximum transmit power of the gNB are $P_k^{\min}=0$ dB and $P_k^{\max}=37$ dB. The azimuth and elevation half-power beamwidths are both set to $65^\circ$. A fleet of five LEOSs flight at altitude 550 km, with carrier frequencies around 24.25 GHz and antenna gains of $G_t^{\mathrm{LEO}}=40$ dBi, is considered.


In Fig. \ref{Fig3}, we illustrate the convergence of the average reward for both the proposed DQN-based algorithm (DQNA) and the benchmark Q Learning-based algorithm (QLA) in the achievable throughput maximization problem. The DQN is trained with a learning rate of 0.001, a batch size of 32 data points, and 20000 iterations. As observed in Fig. \ref{Fig3}, DQN rapily improves the average reward during the first 5000 iterations, reflecting the exploration phase. Afterward, it enters the exploitation phase, resulting in a steady reward increase and reduced fluctuations. In contrast,QLA converges more slowly, with lower average rewards and minimal improvement. Shaded regions represent the 90$\%$ confidence intervals (CIs), showing DQNA’s tighter variability and greater consistency. These results confirm the effectiveness of the proposed DRLA in achieving faster, more stable, and higher-reward convergence than traditional QLA.

Fig. \ref{Fig4} compares the average number of users at different RSRP levels over training iterations. Four RSRP categories are defined: good $>-105$ dBm, fair ($-115$ to $-105$ dBm), poor ($-124$ to $-115$ dBm), and no signal ($<-124$ dBm). In Fig. \ref{fig:4a}, the number of users with good RSRP in the dQNA rapidly increases and stabilizes around 600, showing the DQN agent effectively learns to tune parameters (e.g., antenna tilt and transmit power). The CI narrow after early training, highlighting consistent performance and convergence stability. The number of fair users remains relatively constant, while the number of poor or no signal users drastically reduces, demonstrating the DQNA ability to optimize coverage and signal strength. In contrast, Fig. \ref{fig:4b} shows limited improvement with QLA. More specifically, the number of good RSRP users increases slightly but remains significantly lower than in the DQNA. The fluctuations in the confidence intervals are larger and persist even at higher iteration counts, indicating less stable learning and convergence. Additionally, the QL model fails to significantly reduce the number of poor or no signal users, indicating suboptimal policy learning and network parameter configuration.


\section{Conclusions}
\label{Conclusions}
This paper proposed a novel resilience optimization framework for 6G and beyond ISTNs that leverages LEOSs to ensure continuous service for users in the event of gNB outages. By formulating the problem as a network throughput maximization under constraints of QoS and RSRP requirements, we designed a DQN-based solution that dynamically adapts gNB configurations, including antenna downtilt and transmit power, and seamlessly incorporates LEOS backup when terrestrial resources are exhausted. Extensive simulation results have demonstrated that our proposed approach not only achieves significant improvements in user throughput compared to traditional QL-based method, but also reduces the number of users with poor or no signal while maintaining stable and consistent performance over iterations. The DQN model effectively balances exploration and exploitation, converging to an optimal policy faster and more reliably than QL. This work highlights the potential of DQNA for enhancing network resilience in ISTNs and lays the groundwork for future research on intelligent resource management and energy-efficient operations in 6G and beyond networks. 

\bibliographystyle{IEEEtran}
\bibliography{IEEEfull}

\end{document}